# Physics to the rescue?

Andrea Saltelli, andrea.saltelli@uib.no
Open Evidence Research, Universitat Oberta de Catalunya (UOC), Barcelona (ES),
and Centre for the Study of Science and Humanities, University of Bergen (NO)

Monica Di Fiore, Institute of cognitive sciences and technologies (ISTC),
Consiglio Nazionale delle Ricerche, Roma (IT)

Francesco Spanò, Royal Holloway, University of London (UK)

**Abstract**

A vast body of literature addresses the complex nature of science's reproducibility crisis. In contrast with this perceived complexity, some recent papers from the discipline of physics suggests that irreproducibility does not point to a systemic crisis, but is, on the contrary, a sign that the science system works properly. These works, while acknowledging the difference between physics and other disciplines mired in the reproducibility crisis, hint that all disciplines could learn from Physics. The present work suggests that this optimistic message, when addressed to struggling disciplines, may invite complacency over other relevant dimensions of crisis, delay its solution, and get into the way of a truly joint effort from all disciplines to tackle the important social and environmental predicaments of the present age.

## Context

The present work intersects two relevant issue in today's governance of science. One is science's reproducibility crisis and its perception among different stakeholders. Another is the relation between disciplines, and more broadly between social sciences and humanities on one side and the natural sciences on the other. While copious ink has been poured over these two issues – each of them is conflicted in its own separate way.

As per the crisis in science, a tension exists among those who point to its systemic nature, e.g. (Ravetz 2011; Saltelli and Funtowicz 2017; Saltelli 2018) and those who dismiss it as minor or non-existent, e.g. (Fanelli 2018b; 2018a; Jamieson 2018). Elements of the crisis are seen by some as systemic, and hence of arduous solutions (Smaldino and McElreath 2016; Edwards and Roy 2017). If the crisis will be worse before it can become better (Saltelli 2017), it will remain the occasion for political battles where powerful interests mobilize the crisis e.g. against environmental regulations (Wood and Randall 2018; Saltelli 2018; Oreskes 2018), exacerbating present fights around ecological and health issues. The concern expressed by such considerations prompted efforts to separate the definition of reproducibility (obtaining consistent results when using the same data and analysis methods) and replicability (obtaining consistent results, using new data and similar analysis methods) in order to achieve progress towards the solution of the problem (National Academies of Sciences 2019).

The conflicted relationship between humanities and social sciences on one hand and natural sciences on the other, from the age of Descartes to our days, has been framed by Toulmin as a conflict between the values of Renaissance and those of Enlightenment (Toulmin 1992; 2001). For others, modernity is marred by the persistence of a Cartesian dream of domination and control of nature made possible by mathematics, technology and the sciences (Pereira and Funtowicz 2015; Scoones and Stirling 2020). Culture wars, officially started in the eighties, can be seen alive and well in the present (Frodeman 2020), becoming the occasion for periodic calls for abating disciplinary barriers when new crises, such as the present pandemic, would call for a joint effort from all disciplines in a dialogic relation with one another (Cummings et al. 2020; Waltner-Toews et al. 2020; Saltelli et al. 2020).

A superposition of the narratives and debates about the nature of the present so-called reproducibility crisis and the relation between disciplines offers the occasion for some new insights, which are illustrated in the present work. We focus on the role of physics, and how physics perceives and is perceived in the context of the crisis.

There is a broad agreement among several scholars that, among natural scientists, physicists hold a primacy of virtue when it comes to the diligent execution of experiments and the reproducibility of their results (Fanelli 2010; Nuzzo 2015). A research question tackled in the present work is if the acknowledged prominence and status of the discipline of physics might favour general complacency towards cases of irreproducibility and/or replicability and if this were to be the case, what would be the implication for science, policy and society overall. The discussion moves from a relevant episode which is illustrated next.

## The case

A recent article in Nature Physics suggests that "Lack of reproducibility is not necessarily bad news; it may herald new discoveries and signal scientific progress" (Milton and Possolo 2020). The same piece is taken up in an editorial on Nature titled "Irreproducibility is not a sign of failure, but an inspiration for fresh ideas" (Editorial 2020).

In these two pieces the case for the usefulness of irreproducibility is constructed with examples coming from Physics, its metrology to be precise, and concerning three among the most important universal constants: Plank's $h$, the speed of light $c$, and gravitation's $G$. Here successive determinations – albeit amidst difficulty, gently converge toward more and more precise determinations.

The first piece tends to give emphasis to the following narrative: irreproducibility is the occasion for the metrology community to improve on existing measures. The second paper uses the first to emphasize the narrative that makes irreproducibility a useful part of the scientific process. Both suggest that the practice of metrology can be an example for other disciplines to adopt.

## The critique

Our critique of (Milton and Possolo 2020; Editorial 2020) addresses the hidden/underlying assumptions being made as well as the inferences derived by the authors.

The first underlying assumption is that the sources of non-reproducibility and of non-replicability are overwhelmingly, if not all, technical, and ultimately helpful to advance science. However, any instance of non-replicability generally requires further investigation to ascertain the nature of such sources: some are helpful new "unknown effects or sources of variability", while others range from "simple mistakes to methodological errors to bias and fraud" [NASEM report 2020]). Such sources are additionally different in quality and intensity, depending on the subject being investigated.

The second underlying (but not explicit) assumption that is given credit by (Editorial 2020), building on (Milton and Possolo 2020), is that the examples put forward by (Milton and Possolo 2020) can be extended to other disciplines in terms of convergence towards an accepted value. Such assumption is at odds with intrinsic differences between disciplines: different subjects of investigation lend themselves to different sources of irreproducibility or non-replicability and an approach developed in one subject may not be applicable to all.

In light of the generally different sources of non-reproducibility and non-replicability, it is then generally accepted that the difficulties in reproducibility and replicability are comparably less serious in physics then they are in other scientific fields (Nuzzo 2015): not just in biomedical and social sciences as noted in Nature (Milton and Possolo 2020), or in psychology as noted in Nature Physics (Editorial 2020), but in practically all scientific fields from criminology to nutrition, from organic chemistry to applied economics (Saltelli and Funtowicz 2017). Biomedicine stands out as a field of extreme relevance where the situation of poor reproducibility is mostly evident (Harris 2017; Begley and Ellis 2012; 2012). This is particularly serious when the world is in the midst of a pandemic and when the speed of deployment of new and existing vaccines and cures contends with the speed of insurgence of new variants of the virus. As per reproducibility and replicability, when it comes to large meta-analyses, such as - in medicine, those run under the flagship of Cochrane, the performance of medicine and physics are not all that different. As noted in (Bailey 2017) "medical research uncertainties are generally as well evaluated as those in physics, but physics uncertainty improves more rapidly."

On the other hand it is evident that the examples from metrology presented in (Editorial 2020) offers little insight to orient a research in the design of a medical trial, or in studying the properties of an hydrogeological basin. What about the analysis of wage settings in economics, of regulatory capture in political sciences, or of performative properties of numbers in sociology of quantification? Even remaining in the domain of natural sciences, autopoietic living systems need their own specific methods (Louie 2010). Particle physics is much less affected by disputed values, societally high stakes and the need for urgent decisions (Funtowicz and Ravetz 1993), when compared e.g. with the developments of biomedicine for COVID-19 or cancer.

Thus, the use of examples from one discipline to build a general narrative has a large potential for overlooking systemic differences.

Is the solution hinted to by some member of the community of physics and metrology (Milton and Possolo 2020; Editorial 2020) that all fields become more like physics? Given the limitations of the underlying assumptions, such course would most probably be a problem, rather than a solution, as noted by those who lament the mathematization of economics (Mirowski 1991; Reinert 2000; Drechsler 2000; Romer 2015), the use of statistical rituals in sociology (Gigerenzer and Marewski 2014) or more radically object to a "quantitative-mathematical social science" (Drechsler 2004).

Typically, unhelpful sources of non-reproducibility that are present in other fields are, de facto, absent or reduced in Physics. While the hypothesis that $G$ could change with time is still an active field of research (Uzan 2011), this is rather an exception in physics, while change is the only constant in other disciplines. Modern physics is ordinarily spared serious methodological error such as fighting breast cancer for decades using the wrong cell line ((Harris 2017), p. 99-105). One would need to go back to XVIII century's phlogiston or XIX century's luminoferous aether to find equivalent examples in physics. Physics has no space for the political tensions associated to measuring unemployment, inequality or poverty (Concialdi 2014). The difficulty to ascertain at the beginning of the pandemic even the most basic parameters of the COVID crisis creates a situation of untamed uncertainty which is considerably less common in physics (Daston 2020).

Beyond the evident differences between physics and other fields, the crisis of science has many dimensions, and constraining extant problems under the heading of reproducibility risks serious misrepresenting the issue, offering too reassuring an image of science's health (Saltelli and Funtowicz 2017; Saltelli 2018; Saltelli and Boulanger 2019). A possible taxonomy of science's pains (Ravetz 2016) could include:

− Methodological and epistemological, as in the crisis of reproducibility proper (Ioannidis 2005; Begley and Ellis 2012; Open Science Collaboration (OSC) 2015), and in the convulsions of statistics (Amrhein, Greenland, and McShane 2019; Gelman 2019; Mayo 2018);
− Ethical, in the relationship between the crisis and a perverse system of incentives (Smaldino and McElreath 2016; Edwards and Roy 2017);
− Political, in that the crisis itself become the theatre of conflict between industrial and corporate interests and regulators, whereby the former claim that the science of the latter is irreproducible and should hence be repelled (Saltelli 2018; Oreskes 2018); here the principle of evidence based policy becomes hijacked by those who have the means to produce and enforce their evidence (Ioannidis 2016; Saltelli and Giampietro 2017; Saltelli 2018; Saltelli et al. 2021). Science as a marketplace of ideas favours the actors who have in the market their natural habitat (Mirowski 2011); in this context science becomes itself both victim and perpetrators of processes of regulatory capture fought by corporate interests in the name of 'sound science' (Michaels 2008; Saltelli et al. 2021);

- Ethico-political, in that science is used in the place of religion as a source of legitimization of power (Lyotard 1979; Ravetz 1990; Toulmin 1992);
- Systemic, when science's politicization, mediatization and commoditization are simultaneously at play and the intrinsic code of communication of science as a social system (true/false) is substituted by (profit/loss) or (news/no-news) (Saltelli and Boulanger 2019).

Is irreproducibility "a sign of the scientific method at work" (Milton and Possolo 2020; Editorial 2020) or the most visible symptom of these other dimensions?

While keeping disciplines like physics and metrology and their achievements in due respect, they should not be used to dismiss the seriousness of science's problems nor to deny the reproducibility challenge itself. This would not help all those – from statisticians to medical researchers, engaged to tackle the different aspects of this crisis (Ioannidis 2017; Open Science Collaboration (OSC) 2015; National Academies of Sciences Engineering Medicine 2020).

### From the specific case to the underlying narratives

As discussed, a possible aspect of the case just tackled is complacency. On the one hand the seriousness of the problem of reproducibility is discounted. On the other, some members of the physics community tend to look to their colleagues in fields such as medical or social sciences with what appears to be an attitude of intellectual aloofness: if we can experiment properly, why cannot you?

This is what one might call a minor incident, a sort of disciplinary venial sin. Yet this calls to mind –and perhaps belongs to- a constellation of various aspects of science *qua* science, and of science in relation to society.

- Is complacency, the long wave of the science wars? The Sokal affair in the eighties was meant to 'prove' that social scientists could not detect the vacuity of a paper discussing the hermeneutics of quantum gravity. More recently, the hoax on grievance studies (Pluckrose, Lindsay, and Boghossian 2018), aptly named Sokal-squared, suggested that cultural theory scholars cannot detect the hoax in 'rapist dogs' – one of the articles used in the ruse. If this is the case, these scientists deserve condescending help from more mature disciplines.

- More than between science and humanities (Snow 1959), the conflict is between different scientific tribes. For example chapter nine of Postman's Technopolis (Postman 1992) contains a severe reprimand of the practices of social scientists, while the more recent work of Steven Pinker (Pinker 2018) moves (Chapter 22) an even more unceremonious attack to social sciences and humanities. For (Crowe 1969) there exist two "insular scientific communities - the natural and the social- between which there is very little communication and a great deal of envy, suspicion, disdain, and competition for scarce resources". For Crowe, this insularity gets into the way of science saving the commons, which is the point made in the present work in relation to the urgent challenges facing humanity.

- Should the aspiration to a greater good for science lead the scientific community at large to accept the discounting of the crisis favoured in some quarters as a defensive move to protect science's workings and funding? Or simply to protect one positive (or positivistic) vision of science against one that looks at the present political economy of science (Mirowski 2020)?

- For Clifford D. Conner (Conner 2005), physicists have for a long time belittled disciplines as botany or palaeontology by likening them to stamp collecting. Thus, argues Conner, while the methodologies of biology, anthropology, ecology, psychology and sociology have little in common with that of theoretical physics, physics is placed on a pedestal by the ideology of modern science – an example that all disciplines should strive to emulate. For the same author the *imperialism of physics* – which was boosted by physicists' role in the Second World War - had important consequences for the other disciplines, including foremost the one that science overall must be 'value free', and 'neutral', a move that would force the social sciences away from their critical role and support the *status quo* (Conner 2005). Note that the argument just developed for physics has been made to draw attention to the performative, non-neutral, role of mathematics as an element of education and regimentation (Ernest 2018) – and we say that not to distract from our subject, physics, but to contest – as many before us, ideas of neutrality cherished by a positivistic vision of science.

- Today the challenge to social sciences is more effectively run by evolutionary and cognitive psychologists. If evolutionary psychology can explain all human habits and practices, who needs sociology (Foucart, Horel, and Laurens 2020)? If consumers and voters can be effectively domesticated by cognitive psychology coupled with artificial intelligence (Zuboff 2019), politologists and sociologists are no longer needed.

## Concluding remarks

In some field of physics discrepancies in measurement – known as anomalies, and often related to the validity of the standard model, provide the bread and butter of research at the frontiers of knowledge (Garisto 2020). The pride of such colleagues for their achievements appears justified. After all, we all watch in awe the progress of (particle) physics, its bosons and gravitational waves, its flow of discoveries about the cosmos.

Yet, as discussed in the present work, the 'metrology for all' or 'physics to the rescue' narratives are functional to a reductionist view of science overall.

What some physicists perceive as a rather innocuous form of intellectual snobbism toward the less endowed colleagues in medical or social disciplines is in fact a dangerous affliction of the science system, one which makes science more vulnerable to the convulsions of the present crisis (Saltelli and Funtowicz 2017; Saltelli 2018). While taking in due account physicists' experience in the measurement of fundamental constants, this has relative utility to tackle a situation where science's own quality control apparatus appears challenged. Such a crisis, long predicted (de Solla Price 1963; Ravetz 1971) has surely a vast array of technical solution, but an even broader array of social ones (Smaldino and McElreath 2016; Edwards and Roy 2017). Science is a social activity (Ravetz 1971), and the solution of its problems – linked to those of policy and society – may need a new covenant between science and society in the direction of more participative approaches to deliberation (Waltner-Toews et al. 2020), more than it needs a hierarchical vision of the disciplines of science with physics at the top. Otherwise said, science's role in saving the commons (Crowe 1969) calls for a humbler attitude (Jasanoff 2003), from all disciplines, including physics.